\documentclass[conference]{IEEEtran}
\IEEEoverridecommandlockouts
\usepackage{cite}
\usepackage{amsmath,amssymb,amsfonts}
\usepackage{graphicx}
\usepackage{textcomp}
\usepackage{algorithm}
\usepackage{algpseudocode}
\usepackage{xcolor}
\usepackage[draft]{hyperref}
\usepackage{stmaryrd}

\makeatletter
\renewcommand*{\@textcolor}[3]{%
  \protect\leavevmode
  \begingroup
    \color#1{#2}#3%
  \endgroup
}
\makeatother

\def\BibTeX{{\rm B\kern-.05em{\sc i\kern-.025em b}\kern-.08em
    T\kern-.1667em\lower.7ex\hbox{E}\kern-.125emX}}

\usepackage{titlesec}
\titleformat{\subsubsection}[runin]
  {\normalfont}
  {\thesubsubsection}{0.5em}{}
\titlespacing{\subsubsection}{0pt}{\parskip}{0.5em}

\titlespacing*{\subsection}
{0pt}
{7pt}
{3pt}

\begin{document}

\title{Learning Variable Node Selection for Improved Multi-Round Belief Propagation Decoding\\
\thanks{This work has been funded by the French National Research Agency AI4CODE project (Grant ANR-21-CE25-0006). Part of it was performed using HPC resources from GENCI-IDRIS (Grant 2025-AD011016057)}
}

\author{\IEEEauthorblockN{Ahmad Ismail, Raphaël Le Bidan, Elsa Dupraz, and Charbel Abdel Nour}
\IEEEauthorblockA{\textit{IMT Atlantique, Lab-STICC UMR CNRS 6285, Brest, France} \\
}
}

\maketitle

\begin{abstract}

Error correction at short blocklengths remains a challenge for low-density parity-check (LDPC) codes, as belief propagation (BP) decoding is suboptimal compared to maximum-likelihood decoding (MLD). While BP rarely makes errors, it often fails to converge due to a small number of problematic, erroneous variable nodes (VNs). Multi-round BP (MRBP) decoding improves performance by identifying and perturbing these VNs, enabling BP to succeed in subsequent decoding attempts. However, existing heuristic approaches for VN identification may require a large number of decoding rounds to approach ML performance. In this work, we draw a connection between identifying candidate VNs to perturb in MRBP and estimating channel output errors, a problem previously addressed by syndrome-based neural decoders (SBND). Leveraging this insight, we propose an SBND-inspired neural network architecture that learns to predict which VNs MRBP needs to focus on. Experimental results demonstrate that the proposed learning approach outperforms expert rules from the literature, requiring fewer MRBP decoding attempts to reach near-MLD performance. This makes it a promising lead for improving the decoding of short LDPC codes. 

\end{abstract}


\section{Introduction}

Low-density parity-check (LDPC) codes \cite{Gallager} are powerful error-correcting codes that achieve high efficiency and low decoding complexity when decoded using iterative algorithms such as belief propagation (BP). Due to their strong performance, LDPC codes have been adopted in modern communication standards, including 5G wireless systems. While BP performs effectively at long blocklengths, its performance degrades at short blocklengths, where it remains suboptimal compared to maximum-likelihood decoding (MLD) \cite{Shirva}.

Significant efforts have been made to enhance BP decoding performance for short LDPC codes. A well-established approach, introduced in \cite{Foss}, applies an ordered statistics decoder (OSD) \cite{Foss2} of low order after each BP iteration. OSD leverages soft information from BP by applying a series of low-weight error patterns on the most reliable information set at BP output, and re-encoding the corresponding perturbed variable nodes (VN). This creates multiple candidate codewords at each BP iteration, thereby increasing the chances of finding the most likely one among them, but at the cost of increased complexity. To reduce this overhead, some works \cite{OSD1,OSD2} propose using OSD only as a fallback post-processing step when BP fails to converge.
Another widely studied method is the multi-round BP (MRBP) decoder, also known as augmented BP (ABP) \cite{Varnica}. MRBP builds on the observation that, when BP fails, it rarely converges to an incorrect codeword but instead does not converge at all, typically due to a few problematic erroneous VNs that prevent decoding success. To overcome this, MRBP searches for those problematic VNs by monitoring well-chosen decoding metrics, and once it finds one, perturbs its value to increase the likelihood of successful decoding in subsequent BP iterations. Various MRBP strategies have been proposed in the literature, differing in how they select the VNs to perturb, in the nature of the applied perturbation, and in the way the multiple decoding rounds are scheduled \cite{Gadat, Scholl, Kang, Lee}. While both OSD post-processing and MRBP can boost BP towards MLD, their application usually comes at a prohibitive computational cost (very large number of re-encoding or decoding attempts).

Learning has recently emerged as a promising approach to further enhance the performance of BP decoding. The earliest and most natural approach to supplement BP with learning is neural BP (NBP)~\cite{NBP}, which models BP iterations as layers of a neural network (NN), incorporating trainable weights that are optimized through learning. However, in the context of LDPC codes, NBP has demonstrated so far limited performance improvement compared to standard BP, unless the number of iterations is very low. Therefore research efforts have progressively shifted towards leveraging learning to break the complexity bottleneck in improved BP-based decoders. One such line of work involves augmenting OSD post-processing with lightweight NNs \cite{Rosseel2, Boosting}. These NNs prove more effective at identifying the most reliable VNs at BP output, thereby reducing the number of re-encodings and enabling a reduction in the OSD maximum reprocessing order, without compromising performance. Another approach consists in supplementing MRBP with a learning architecture in charge of predicting the VNs that hinder BP decoding, as accurate predictions lead to fewer additional decoding rounds. For instance, \cite{Lee2} proposes to replace the expert rules devised for this purpose with a multi-layer perceptron (MLP). Yet this work does not provide conclusive evidence that a NN can be consistently more accurate at this task than a well-designed heuristic. Another related work is \cite{NBPD} which combines neural BP with an MRBP decoder (referred to as list-based decimation) followed by a learned decimation stage where a simple NN is used to calculate a perturbation for each VN in an automated manner. Learned decimation is shown to boost performance, but cannot work effectively without the classical MRBP stage that precedes it. Therefore, the scheme \cite{NBPD} inherits the same complexity issues as MRBP. 

In this work, we focus on pushing further the potential of learning to enhance MRBP decoding. Expanding on~\cite{Lee2}, our goal is to demonstrate that, with a proper learning architecture, NNs can accurately estimate which VNs should be perturbed, thereby outperforming expert rules and bringing MRBP  performance closer to that of MLD, with fewer decoding attempts. Such a scheme could also be supplemented with the learned decimation stage of \cite{NBPD} for further performance gains.

To this end, we relate the task of identifying the VNs to perturb to the problem of estimating the errors at the channel output, which is precisely the problem addressed by syndrome-based neural decoders (SBND)~\cite{Bennatan}. Since the set of VNs to perturb forms a subset of these errors, and given the demonstrated ability of SBND to accurately estimate channel errors, we propose to take advantage of these models and show how to adapt them to the VN selection problem in MRBP.

The remainder of the paper is structured as follows. Section~\ref{MRBP} provides an overview of MRBP decoding, emphasizing its key features. Section \ref{NN-MRBP} introduces the NN-aided MRBP approach as initially proposed in \cite{Lee2}. In section \ref{Prop Implementation}, we first establish the link between MRBP and SBND, and then show how to leverage SBND to refine the NN setup of \cite{Lee2}, with the proper adaptations. Experimental results are presented and discussed in section \ref{Results}. Finally, section \ref{Conc} concludes the paper.

\section{MRBP Framework}\label{MRBP}

\subsection{BP Decoding}

 Let $\mathcal{C}$ be an $(n, k)$ binary LDPC code of length $n$ and dimension $k$. Let $\mathbf{H}$ be its parity-check matrix with size $m\times n$, where $m$ is the number of parity-check equations. Assume a codeword $\mathbf{c}=(c_1, \ldots, c_n)$ is modulated into the binary-phase-shift keying (BPSK) sequence $\mathbf{x}=(x_1, \ldots, x_n)$, using the mapping $\mathbf{x} = (-1)^\mathbf{c}$, and transmitted over a binary-input additive white gaussian noise (BI-AWGN) channel with noise variance $\sigma^2 = \frac{N_0}{2}$. Let $\mathbf{y}=(y_1, \ldots, y_n)$ be the received word. Denote by $L^{\text{ch}}_{i}$ the channel output LLR  for the code bit $i$:

\begin{equation}
    L^{\text{ch}}_{i} = \log \left( \frac{P(y_i  | c_i= 0)}{P(y_i| c_i = 1)} \right) = \frac{2 y_i}{\sigma^2}.
\end{equation}

Let $\mathbf{z}$ the hard decision on the received word $\mathbf{y}$, where $z_i = \mathbb{I}\{ y_i \leq 0 \}$ for each code bit $i$, and $\mathbf{s^{ch}}=\mathbf{z}\mathbf{H^T}$ be the corresponding syndrome. If $\mathbf{s^{ch}} \neq 0$, BP decoding is initiated.

BP operates iteratively, exchanging soft information in LLR form between VNs and check nodes (CN) in the Tanner graph. At each iteration, VNs send extrinsic messages to CNs, and CNs propagate updated extrinsic messages back to VNs. The a posteriori probability (APP) LLR, denoted by $\mathbf{L^{app}}$, is then computed, and the estimated code bits $\hat{c_i}$ are obtained as $\mathbb{I}\{ L_i^{\text{app}} \leq 0 \}$. BP terminates either when $\mathbf{\hat{c}H^T}=0$ or a predefined maximum number of iterations $l_0$ is reached.

\subsection{MRBP Decoding}

In scenarios where the initial BP decoding attempt fails, the rationale underlying the application of MRBP is to construct a set of VNs that, with high probability, includes the one or few VNs primarily responsible for the decoding failure. Perturbations are then applied to these VNs, followed by another round of BP decoding with $l_{1} \leq l_0$ iterations for each perturbation. Designing an MRBP decoder requires tackling three key questions: How to select the candidate VNs to be perturbed? How to perform the perturbations? How to organize the subsequent decoding attempts?  

\subsubsection*{\textit{1) How to select the VNs to be perturbed?}}

Accurate selection of the VNs to perturb is crucial for the performance of MRBP, as an informed choice can enable successful BP decoding with reduced average number of decoding attempts. A common approach consists in ranking VNs based on an appropriate reliability metric. Hereafter, a VN is considered \emph{less reliable} than another if it is \emph{more likely to have contributed to the initial decoding failure}. These metrics form the basis of what we refer to as expert rules, and can be classified into pre- and post-BP metrics. The channel LLR magnitude ($\mathbf{|L^{ch}|}$) serves as an example of pre-BP metric, wherein VNs with low $\mathbf{|L^{ch}|}$ are classified as unreliable \cite{Varnica, Scholl}. Similarly, the magnitude of the APP LLR  ($|\mathbf{L^{app}}|$) constitutes a post-BP metric, wherein VNs with low $|\mathbf{L^{app}}|$ are deemed unreliable. Other post-BP metrics focus on tracking VN oscillations across decoding iterations \cite{Gadat,Kang,Lee}, as frequent sign changes are generally considered as a more accurate indicator of unreliability than LLR magnitudes. A recent example is the nSMEA method \cite{Lee}, which quantifies oscillations by counting the \underline{n}umber of \underline{S}ign \underline{M}ismatches between \underline{E}xtrinsic and \underline{A}PP LLR messages associated with each VN.

\subsubsection*{\textit{2) How to perform the perturbations?}}

Several strategies have been proposed for perturbing the identified unreliable VNs. Here, following \cite{Lee2}, we adopt single-bit perturbations. Specifically, let $\pi(.)$ be the permutation that orders the VNs from least to most reliable based on the chosen reliability metric. Then, $T$ distinct perturbed versions $\big\{\mathbf{L^{ch,t}}\big\}_{t \in \llbracket 1,T\rrbracket}$ of the received channel LLR vector $\mathbf{L^{ch}}$ are generated by flipping and saturating the channel LLR of the $t$-th least reliable VN, while keeping all other channel LLR values unchanged:
\begin{equation}
    L_i^{ch,t}  = \begin{cases} L_i^{ch} & i \neq \pi(t) \\ -\infty \cdot \text{sign}(L_i^{ch}) & i = \pi(t) \end{cases}\;,~t \in \{1,\ldots, T\}
\end{equation}

\subsubsection*{\textit{3) How to organize the subsequent decoding attempts?}}\label{Dec Procc}

\begin{figure}[t!]
    \centering
    \includegraphics[width=\linewidth]{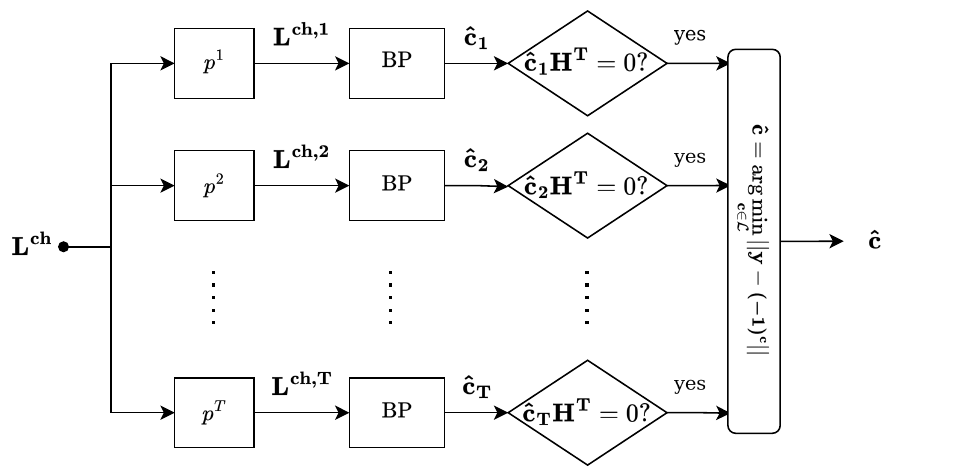}
    \caption{Organization of the augmented decoding rounds within the MRBP framework ($\mathcal{P}_t$ = single-bit perturbation applied on the $t$-th least reliable VN).}
    \label{fig:MRBP-Diagram}
\end{figure}

Decoding attempts can be performed either sequentially or in parallel. In this work, we adopt the parallel execution strategy depicted in Fig. \ref{fig:MRBP-Diagram}, which offers performance comparable to its sequential counterpart \cite{Varnica} with the additional advantage of reduced latency. Upon completion of all decoding attempts, candidate codewords  are collected into a list $\mathcal{L}$. The decoder decision $\mathbf{\hat{c}}$ is selected as the most-likely codeword in this list: 
\begin{equation} \mathbf{\hat{c}} = \arg\min_{\mathbf{c} \in \mathcal{L}} ||\mathbf{y} - \mathbf{\small (-1)^{c}}||. \end{equation}

\section{NN-aided MRBP}\label{NN-MRBP}

\subsection{Principle}

As outlined in Section~\ref{MRBP}, accurately identifying the least reliable VNs to perturb is critical to MRBP performance. While conventional methods rely on expert rules, the approach in \cite{Lee2} leverages an NN model for this task. Given the received word $\mathbf{y} = (y_1, \ldots, y_n)$, the NN outputs a vector $\mathbf{\hat{b}} = (\hat{b}_1, \ldots, \hat{b}_n)$, where each \( \hat{b}_i \in [0,1] \) represents an estimate of the probability  that perturbing the channel LLR $L^{\text{ch}}_i$ of VN $i$ will lead to successful BP decoding. Based on this output, the $T$ VNs with the largest $\hat{b}_i$ values are selected as the least reliable VNs and targeted for perturbation.

\subsection{Architecture}

In \cite{Lee2} an MLP is used to predict which VNs to perturb. The MLP is made of three fully connected layers: two hidden layers with dimensions $2n$ and $n$, respectively, followed by an output layer of size $n$, with Rectified Linear Unit (ReLU) activation functions in-between.

\subsection{Training}\label{Train}

In \cite{Lee2}, the training dataset is generated by transmitting random codewords over a BI-AWGN channel at various signal-to-noise ratio (SNR) levels, followed by BP decoding with $l_0$ iterations. The received words that result in BP decoding failures are retained for further processing. For each such word, and for each VN $i \in \{1, \cdots, n\}$, a single-bit perturbation is applied to $L_i^{\text{ch}}$, and BP decoding is executed again with $l_1$ iterations. This process yields a binary target vector $\mathbf{b} = (b_1, \ldots, b_n)$, where $b_i = 1$ if the perturbation of VN $i$ leads to successful BP decoding in the second attempt, and $b_i = 0$ otherwise. The final dataset comprises the $N_{tr}$ input-output pairs:
\begin{equation}
    \mathcal{D}_1 = \Big\{\mathbf{y}_j,\mathbf{b}_j\Big\}, ~~~ j \in  \{1,\cdots, N_{tr}\}.
\end{equation}

The goal of the training process is to optimize the NN parameters by minimizing the discrepancy between the predicted probability vector $\hat{\mathbf{b}}$ and the true label vector $\mathbf{b}$. To this end, \cite{Lee2} adopts the mean squared error (MSE) as the loss function.

\section{Proposed Learned MRBP Approach}\label{Prop Implementation}

The results in \cite{Lee2} do not conclusively demonstrate that NNs outperform well-designed expert rules in selecting unreliable VNs for perturbation. In particular, it is unclear whether the very small gains reported for certain codes are due to the selected architecture (a simple MLP), to an improper training setup (\cite{Lee2} reports that augmenting the model input can help), or to more fundamental limitations of NNs in this context. These uncertainties motivate further investigation. 

We observe that the task of identifying the small set of erroneous bits responsible for BP decoding failures is closely related to the more challenging problem of estimating the full error pattern introduced by the channel, as these critical bits form a subset of the latter. 
Syndrome-based neural decoding (SBND), introduced in \cite{Bennatan}, has been shown to effectively estimate the most likely channel error pattern and achieve near-MLD performance for certain short  codes.
Motivated by these results, we propose to revisit the approach in \cite{Lee2} by integrating SBND to enhance the prediction of VNs to perturb. The following subsections describe the modifications made to both the network architecture and training procedure, relative to the original NN-aided MRBP framework of \cite{Lee2}.

\subsection{A Different Model Architecture}

The fact that the NN-aided MRBP scheme in \cite{Lee2} relies on a very simple, shallow MLP to make its predictions could be a limiting factor for the performance. This is further supported by the findings in \cite{Bennatan}, which reveal that stacked Gated Recurrent Unit (GRU) networks are significantly more effective at inferring ML error patterns than MLPs. Building on this insight, we propose replacing the original MLP with a stacked GRU in the architecture of the NN-aided MRBP decoder. A stacked GRU consists of multiple layers of GRU cells, which communicate through hidden states across both time steps and layers. The hidden state dimension serves as a key hyperparameter that governs the model’s representational capacity. As we demonstrate in the following sections, this architectural shift proves to be central in improving the overall accuracy of the predictions.

\subsection{A Different Model Input and Training Recipe}

It is well established that the pair $(\mathbf{|L^{ch}|}, \mathbf{s^{ch}})$ forms a sufficient statistic for MLD, see e.g. \cite{Bennatan}. Given the demonstrated effectiveness of this input representation in the SBND framework, we adopt it as the input for our NN architecture. Compared to \cite{Lee2} which relies solely on channel LLRs, the inclusion of the syndrome introduces additional structure and error signatures that can enhance the model's ability to identify unreliable VNs. Furthermore, this representation removes any explicit dependence on the transmitted codeword, thereby simplifying training: BP failures can now be generated using simulations of the all-zero codeword without loss of generality. The output labels are obtained following the same procedure described in Section \ref{Train}. As a result, our dataset consists of the $N_{tr}$ input-output pairs:
\begin{equation}
    \mathcal{D}_2 = \Big\{ \big (|\mathbf{L^{ch}}|, \mathbf{s^{ch}} \big )_j \space , \space \mathbf{b}_j\Big\}, ~~~ j \in  \{1,\cdots, N_{tr}\}.
\end{equation}
The drawback of incorporating the syndrome at the model's input is that the complexity of the latter is now a function of the sum of the code length and the number of parity bits

In contrast to \cite{Lee2}, we formulate the task as a binary classification problem, where the model predicts, for each VN, whether it should be perturbed to enable successful decoding.  Accordingly, the model is trained to minimize the binary cross-entropy (BCE) loss:
\begin{equation}\label{loss}
    \mathcal{L_{BCE}}\big(\mathbf{b},\mathbf{\hat{b}}\big) = - \frac{1}{n} \sum_{i=1}^{n} \Big( b_i \log \hat{b}_i + (1 - b_i) \log (1 - \hat{b}_i) \Big).
\end{equation}

Furthermore, we have noticed that, for most codes, generating data at one carefully selected SNR point is sufficient for the model to generalize well over a large range of SNR values.

\section{Experimental Results}\label{Results}

Experiments are conducted on a $(96, 48)$ Quasi-Cyclic irregular LDPC code from \cite{channelcodes}, which also provides reference ML performance for this code. The dataset consists of 60M BP failure examples collected at 3 dB, using $l_0 = l_1 = 20$ iterations. All models are implemented in PyTorch and trained with the BCE loss \eqref{loss} for 250 epochs. Training employs the Adam optimizer with a learning rate of $10^{-4}$ along with a Reduce-on-Plateau scheduler monitored by the validation loss. 

\subsection{Choice of input representation}

To assess the impact of changing the input representation compared to the original approach of \cite{Lee2}, we first train the MLP architecture from  \cite{Lee2}, hereafter referred to as \textbf{MLPA}, either with $\mathbf{y}$ as input (dataset $\mathcal{D}_1$) or with the pair $(\mathbf{|L^{ch}|}, \mathbf{s^{ch}})$ (dataset $\mathcal{D}_2$). This results in two slightly different models:

\begin{itemize}
    \item \textbf{MLPA-$\mathcal{D}_1$}: Trained on $\mathcal{D}_1$, this model replicates the three-layer architecture proposed in~\cite{Lee2}, with two hidden layers of size $2n$ and $n$, respectively, followed by an output layer of size $n$, resulting in 46k trainable parameters.

    \item \textbf{MLPA-$\mathcal{D}_2$}: Trained on $\mathcal{D}_2$, this model has the same architecture as \textbf{MLPA-$\mathcal{D}_1$}, but with an input size of $2n-k=144$ instead of $96$. Accordingly, the size of the first hidden layer is set to 155 instead of $2n = 192$, to maintain a total of 46k trainable parameters, thereby enabling a fair comparison between the two models.
\end{itemize}
In both models, each hidden layer is followed by a ReLU activation and a dropout of 0.1. A sigmoid is applied after the output layer to turn logits into probability estimates $\mathbf{\hat{b}}$.

\begin{figure}
    \centering
    \includegraphics[width=0.98\linewidth]{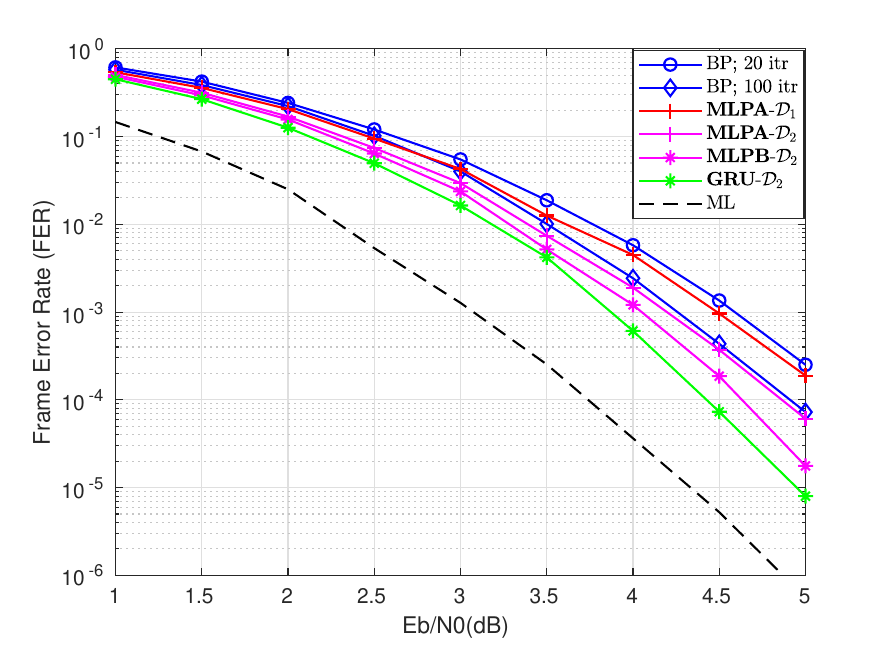}
    \caption{FER performance as function of $E_b/N_0$ for $T$ = 5 MRBP decoding attempts, using different models to select the unreliable VNs to perturb.}
    \label{fig:Ablation-Diagram}
\end{figure}

Fig.~\ref{fig:Ablation-Diagram} compares their performance under the MRBP decoding setup outlined in Section~\ref{Dec Procc}. With $T = 5$ decoding attempts, the five VNs marked by the models as least reliable are flipped and saturated before the second decoding round. Distinct codewords are collected, and the most-likely candidate is returned. The figure shows that training on $\mathcal{D}_2$ significantly improves FER performance compared to $\mathcal{D}_1$, confirming that providing the combination of $|\mathbf{L^{\text{ch}}}|$ and $\mathbf{s}^{\text{ch}}$ as input does contribute to increasing the model’s accuracy.

\subsection{Choice of the architecture model}

Having established that the pair $(\mathbf{|L^{ch}|}, \mathbf{s^{ch}})$ provides a more effective input representation, we now explore whether a GRU-based architecture can outperform an MLP by training the following two models on dataset $\mathcal{D}_2$:

\begin{itemize}
    \item \textbf{GRU-$\mathcal{D}_2$}: A 20.6M-parameter 5-layer stacked GRU with hidden size set to $6(2n-k)$ following~\cite{Bennatan}. It runs for 5 time steps, with the final hidden state passed through a fully connected layer of size $n$ followed by a sigmoid activation. Dropout of 0.1 is applied within the GRU.

    \item \textbf{MLPB-$\mathcal{D}_2$}: A 20.6M-parameter best-effort custom MLP featuring seven hidden layers of 1835 neurons each, with 0.1 dropout and ReLU activation in-between, and an output layer of size $n$ followed by sigmoid activation.
\end{itemize}

As illustrated in Fig.~\ref{fig:Ablation-Diagram}, with an equivalent number of parameters, the GRU model demonstrates superior accuracy compared to a best-effort MLP in identifying the problematic VNs for perturbation, aligning with the results in \cite{Bennatan}. Notably, both models outperform BP decoding with 100 iterations, underscoring the efficacy of learned MRBP in enhancing BP performance. In addition, with the parallel decoding setup of Fig. \ref{fig:MRBP-Diagram}, ignoring model inference time, the worst-case MRBP latency does not exceed $l_0+l_1=40$ iterations of BP decoding.

\subsection{Identifying VNs to perturb: Expert vs learning-based rules}
We compare our best-performing model, \textbf{GRU-$\mathcal{D}_2$}, with the strongest expert rule we are aware of, namely nSMEA \cite{Lee} (denoted by SMV in the original paper). This post-BP metric has been shown to be one of the most effective at identifying the least reliable VNs to perturb, making it a strong baseline for comparison. The least reliable VNs are the ones having the largest cumulated nSMEA values across iterations. To ensure a fair comparison with our NN-aided selection rule, we adopt the same decoding strategy as detailed in \ref{Dec Procc} with the sole modification of switching the VN selection rule from a learning-based approach to nSMEA.

\begin{figure}[t!]
    \centering
    \includegraphics[width=0.98\linewidth]{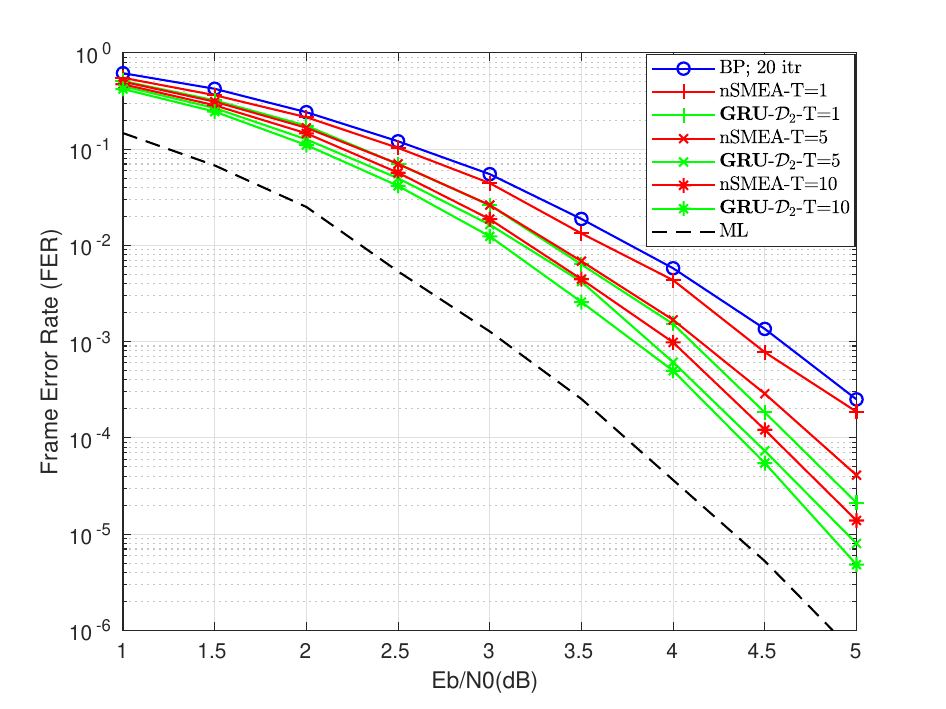}
    \caption{FER performance as function of $E_b/N_0$ for MRBP decoding using the model \textbf{GRU-$\mathcal{D}_2$} or the nSMEA expert rule for unreliable VNs selection.}
    \label{fig:GRUvsExpert-Diagram}
\end{figure}

As depicted in Fig.~\ref{fig:GRUvsExpert-Diagram}, our trained model significantly outperforms the expert nSMEA rule with just one perturbation where only the least reliable VN is perturbed ($T = 1$). This superiority is maintained as more perturbations are tested in parallel, indicating that learning-based methods can more accurately predict which bits to perturb in MRBP decoding. On the other hand, running a 20.6M-parameter DNN for that task is significantly more complex than tracking oscillations in the messages exchanged during iterative decoding.

\subsection{Performance comparison with other selected decoders}

\begin{figure}[t!]
    \centering
    \includegraphics[width=0.98\linewidth]{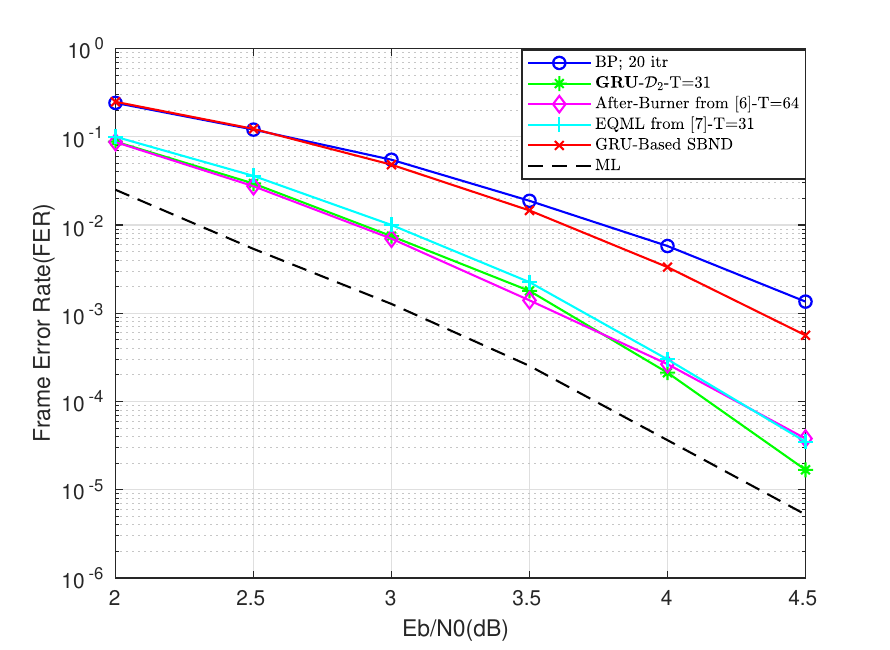}
    \caption{FER performance as function of $E_b/N_0$ for MRBP with \textbf{GRU-$\mathcal{D}_2$} model compared to other selected reference decoders on the $(96,48)$ code.}
    \label{fig:GRUvsSTOAt-Diagram}
\end{figure}

We compare our \textbf{GRU-$\mathcal{D}_2$} learning-based MRBP decoder on the $(96,48)$ code against two reference decoders from \cite{Scholl} and \cite{Kang}. which use different MRBP variations (expert rule and scheduling, see \cite{Scholl} and \cite{Kang} for implementation details). As shown in Fig.~\ref{fig:GRUvsSTOAt-Diagram}, our model matches the performance of \cite{Scholl} with half the number of perturbations and surpasses \cite{Kang} with  the same number. These results demonstrate that the proposed learning-based approach can both reduce the number of decoding attempts and enhance MRBP performance, nearing MLD levels, compared to other existing solutions. 

In the same figure, we also provide results for a best-effort GRU-based SBND model with 39M parameters, trained on the $(96,48)$ code in the manner of \cite{Bennatan}. Compared to this standalone neural decoder, our \textbf{GRU-$\mathcal{D}_2$} learned MRBP decoder achieves superior performance with only half the number of parameters, confirming that learning is more effective at supplementing existing decoders than at replacing them.

\section{Conclusion}\label{Conc}

We have explored the integration of syndrome-based neural decoder models to enhance MRBP's effectiveness in identifying problematic bits that hinder BP decoding success. Our findings demonstrate that learning-based approaches significantly reduce the total number of decoding attempts required to achieve successful decoding, thereby approaching the performance of MLD. However, current models have an impractically large number of parameters, prompting the need for more efficient and scalable dedicated learning architectures.

\bibliographystyle{IEEEtran}
\bibliography{Biblio}

\end{document}